\documentclass[longbibliography, aps, prx, amsmath, amssymb, amsfonts, twocolumn,superscriptaddress, footinbib]{revtex4-2}
\usepackage{amssymb}
\usepackage{graphicx}% Include figure files
\usepackage{dcolumn}% Align table columns on decimal point
\usepackage{bm}% bold math
\usepackage{bbold}
\usepackage{textcomp}
\usepackage{amsmath}
\usepackage[titletoc]{appendix}

\usepackage{tabularx}
\usepackage{verbatim}
\usepackage{hyperref}
\hypersetup{                    
	colorlinks,
	citecolor=blue,
	filecolor=blue,
	linkcolor=blue,
	urlcolor=blue
}

\newtheorem{Def}{Definition}

\newcounter{RomanNumber}

\begin{document}
	\title{Efficient Quantum Digital Signatures over Long Distances with Likely Bit Strings}
	
	\author{Ji-Qian Qin}
	\affiliation{ State Key Laboratory of Low Dimensional Quantum Physics, Department of Physics, \\
		Tsinghua University, Beijing 100084, China}
	\author{Zong-Wen Yu}
	\affiliation{ Data Communication Science and Technology Research Institute, \\Beijing 100191, China}
	\author{Xiang-Bin Wang}
	\email{ xbwang@mail.tsinghua.edu.cn}
	\affiliation{ State Key Laboratory of Low Dimensional Quantum Physics, Department of Physics, \\ Tsinghua University, Beijing 100084, China}
	\affiliation{ Jinan Institute of Quantum technology, SAICT, Jinan 250101, China}
	\affiliation{ Shanghai Branch, CAS Center for Excellence and Synergetic Innovation Center in Quantum Information and Quantum Physics, University of Science and Technology of China, Shanghai 201315, China}
	\affiliation{ Shenzhen Institute for Quantum Science and Engineering, and Physics Department, Southern University of Science and Technology, Shenzhen 518055, China}
	\affiliation{ Frontier Science Center for Quantum Information, Beijing 100084, China}

	\begin{abstract}
		
		Quantum digital signatures (QDSs) can provide information-theoretic security of messages against forgery and repudiation. Compared with previous QDS protocols that focus on signing one-bit messages, hash function-based QDS protocols can save quantum resources and are able to sign messages of arbitrary length. Using the idea of likely bit strings, we propose an efficient QDS protocol with hash functions over long distances. Our method of likely bit strings can be applied to any quantum key distribution-based QDS protocol to significantly improve the signature rate and dramatically increase the secure signature distance of QDS protocols. In order to save
		computing resources, we propose an improved method where Alice participates in the verification process of Bob and Charlie. This eliminates the computational complexity relating to the huge number of all likely strings. We demonstrate the advantages of our method and our improved method with the example of sending-or-not-sending QDS. Under typical parameters, both our method and our improved method can improve the signature rate by more than $100$ times and increase the signature distance by about $150$ km compared with hash function-based QDS protocols without likely bit strings.
		
	\end{abstract}
	
	\maketitle
	\section{Introduction}
	
	Digital signature~\cite{Diffie1976New} is a useful technique for guaranteeing the security of message transfer among the signer and receivers, i.e., it can provide security of the message against forging and repudiation. Forgery means that a receiver accepts a message from a forger rather than from the signer and repudiation means that the signer can repudiate his or her signature. Different from classical digital signatures, quantum digital signatures (QDSs)~\cite{gottesman2001quantum} can provide information-theoretic security that does not depend on computational complexity.
	
	With the removal of the two assumptions of quantum memory and secure quantum channels~\cite{dunjko2014quantum,yin2016practicalQDS,amiri2016secure}, QDS becomes practicable~\cite{donaldson2016experimental,collins2016experimental,puthoor2016measurement,roberts2017experimental,yin2017experimental102,yin2017experimental,zhang2018proof,an2019practical}. 
	Recently, there have been many breakthroughs ~\cite{BraunsteinMDI2012,lo2012measurement,zhou2016making,caoMDI2023,LuTF2018,wang2018twin,wang2022twin,liu2023experimental,fan2022robust}.
	in quantum key distribution (QKD). Based on mature QKD devices, one can sign one-bit messages with bit strings generated by quantum communication of QKD~\cite{amiri2016secure}. 
	The sending-or-not-sending (SNS) protocol~\cite{wang2018twin} can tolerate large misalignment error, while retaining measurement device independent (MDI) security and a high key rate of twin-field (TF) QKD~\cite{LuTF2018} on the scale of the square root of channel transmittance.
	Naturally, 
	SNS QDS
	has higher efficiency over long distances ~\cite{zhang2021twin,qin2022quantum} compared with BB84 (the quantum key distribution scheme developed by Bennett and Brassard in 1984) QDS and MDI QDS. 
	
	Compared with one-bit messages, QDS of multibit messages is more practicable. 
	It would be inefficient to simply iterate a QDS based on one-bit messages to sign multibit messages. In classical digital signature, one can use the hash function method to sign multibit messages~\cite{menezes1997handbook,krawczyk1994lfsr} with classical secure keys.
	Naturally, if we replace the classical secure keys here by the secure final keys from QKD, one can have a type of QDS with quantum security. In this type of implementation, the secure distance of QDS is just that of QKD. 
	To further improve the secure distance and signature rate of QDS, here, using the idea of likely bit strings~\cite{amiri2016secure}, we propose to use the raw key without any 
	error correction or private amplification to sign multibit messages. Note that the method given in Ref.~\cite{amiri2016secure} is for one-bit messages, whereas we design a new hash function-based QDS protocol to sign messages of arbitrary length. 
	Moreover, different from Refs.~\cite{amiri2018efficient,Yin2022QDShash,kiktenko2022practical}, which use final keys (with error correction and private amplification) as well as hash functions to sign and verify messages of arbitrary length, using the idea of likely bit strings, we perform the QDS protocol with raw keys as well as hash functions. We then propose an improved method where Alice participates in the verification process of Bob and Charlie. This improved method removes the computational complexity such as listing the huge number of hashing values of all likely strings.

	%%%%%%%%%%%%%%%%%%%%%%%%%%%%%%%%%%%%%%%%%%%%%%%%%%%%%%%%%%%%%%%%%%%%%%%%%%%%%%%%
	
	\section{quantum digital signature with likely bit strings}\label{sec2}
	
\begin{figure}
		\centering
		\includegraphics[scale=0.58]{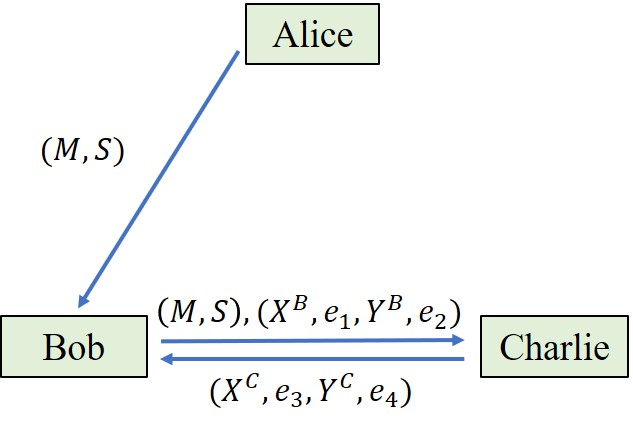}
		\caption{Our efficient QDS over long distances with likely bit strings. Alice is a signer and sends the message and the corresponding signature $(M,S)$ to a receiver, Bob. Bob forwards $(M,S)$ to another receiver Charlie, along with the bit strings and the corresponding bit-flip error rates $(X^B,e_1,Y^B,e_2)$.  Then, Charlie sends the bit strings and the corresponding bit-flip error rates $(X^C,e_3,Y^C,e_4)$ to Bob. Bob (Charlie) determines whether to accept the message $M$ based on the result of the comparison function $\mathbb{C}^B$~\eqref{equ_comB} ($\mathbb{C}^C$~\eqref{equ_com}).}\label{picture_1}
	\end{figure}

	Consider a QDS protocol with one signer and two receivers. As shown in Fig.~\ref{picture_1}, Alice is a signer and Bob and Charlie are two receivers. 
	In order to avoid misunderstanding, we emphasize that, different from the usual use of Charlie for an unreliable relay's name in MDI QKD, here we use Charlie for one receiver's name.
	Alice signs the $m$-bit message $M$, and Bob (Charlie) decides whether to accept $M$ based on the result of comparison between the actual digests and the expected digests, where the generation of digests requires hash functions~\cite{krawczyk1994lfsr} derived from associated bit strings among the signer and receivers. In our QDS protocol, instead of using final keys of QKD to generate hash functions, we use raw keys to generate hash functions. 
	
	Alice and Bob (Charlie) perform a key-generating protocol (KGP) to generate the associated bit strings $X^B_A$ and $X^B$ ($X^C_A$ and $X^C$), where $X^B_A$ ($X^C_A$) belongs to Alice and $X^B$ ($X^C$) belongs to Bob (Charlie). The KGP is the quantum communication of QKD, and there are many kinds, including BB84~\cite{BENNETT20147}, MDI~\cite{lo2012measurement,BraunsteinMDI2012,zhou2016making,jiang2021higher}, SNS TF~\cite{LuTF2018,wang2018twin}, etc. After performing the KGP, Alice, Bob, and Charlie each have bit strings $\{X^A,X^B,X^C\}$ with $n$ bits, where $X^A=X^B_A\oplus X^C_A$, and $\oplus$ denotes bit addition modulo $2$. They use the same process to generate another set of associated bit strings $\{Y^A,Y^B,Y^C\}$ with $2n$ bits, where $Y^A=Y^B_A\oplus Y^C_A$ and $Y^B_A$, $Y^B$ ($Y^C_A$, $Y^C$) are generated in KGP between Alice and Bob (Charlie). We denote the bit-flip error rate between bit strings $X_A^B$ and $X^B$, $Y_A^B$ and $Y^B$, $X_A^C$ and $X^C$, and $Y_A^C$ and $Y^C$ as $e_1$, $e_2$, $e_3$, and $e_4$, respectively. 
	
	\subsection{Signing stage}
	
	Alice uses local $n$-bit quantum random numbers $p^A$ to randomly generate an irreducible polynomial $p$ of degree $n$~\cite{menezes1997handbook,Yin2022QDShash}. After that, based on the irreducible polynomial and the local bit string $X^A$, Alice generates a linear feedback shift register (LFSR)-based Toeplitz matrix~\cite{krawczyk1994lfsr} $H^A$ with $n$ rows and $m$ columns, where $n$ and $m$ are the lengths of $X^A$ and message $M$, respectively.
	Alice acts the matrix $H^{A}$ on the message $M$ to generate the hash value $h$, 
	
	\begin{equation}
		H^{A}\cdot M=h,
	\end{equation}
	where $h$ is a vector of $n$ bits. It should be noted that in practice, the LFSR-based Toeplitz matrix and the hash value can be computed simultaneously in order to reduce the time consumption. In order to make it easier to understand, we present it in a step-by-step manner in this work.
	
	Based on the above steps, Alice generates a digest $D$ of message $M$, 
	
	\begin{equation}
		D = (h,p^A),
	\end{equation}
	where $D$ is a $2n$-bit string  consisting of $h$ with $n$ bits and $p^A$ with $n$ bits.
	
	Next, Alice encrypts the digest $D$ with a bit string $Y^A$ with $2n$ bits to generate the signature $S$ of the message $M$,
	
	\begin{equation}
		S = D\oplus Y^A.
	\end{equation}
	
	At the end of the signing stage, Alice sends the message along with the corresponding signature $(M,S)$ to a receiver for verification.
	
	\subsection{Verification stage}
	
	The receivers Bob and Charlie generate a series of likely bit strings based on the local bit strings and the bit-flip error rates. They use these likely bit strings for verification.
	
	\subsubsection{Verification of Bob}
	As shown in Fig.~\ref{picture_1},
	when Bob receives the message and the corresponding signature $(M,S)$ from Alice, he forwards them to Charlie and sends the local bit strings and the corresponding bit-flip error rates $(X^B,e_1,Y^B,e_2)$ to Charlie. Then, Charlie sends the local bit string and the corresponding bit-flip error rates $(X^C,e_3,Y^C,e_4)$  to Bob.

	Based on the bit strings $X^B$, $X^C$ and the corresponding bit-flip error rates $e_1$, $e_3$, Bob generates a series of likely bit strings $\{K_i^{XB}\}$ locally,
	
	\begin{equation}
		\{K_i^{XB}\} = f(X^B\oplus X^C,e_1,e_3),
	\end{equation}
	where $i=1,2,\dots,n_x$, and $n_x$ is the number of likely bit strings,
	
	\begin{equation}
		\begin{split}
			& n_x = 2^{nH_2(E_x)}.\\
		\end{split}
	\end{equation}
	in which $H_2(.)$ is the binary Shannon entropy. The function $f(X^B\oplus X^C,e_1,e_3)$ means that
	based on the bit string $X^B\oplus X^C$ and the corresponding bit-flip error rate $E_x$, which has a maximum value of $(e_1+e_3)$, Bob can generate likely bit strings $\{K_i^{XB}\}$ relative to $X^A$, i.e., $X^A\in \{K_i^{XB}\}$. 
	
	Similarly, Bob can generate a series of likely bit strings $\{K^{YB}_j\}$ relative to $Y^A$,
	
	\begin{equation}
		\{K^{YB}_j\} = f(Y^B\oplus Y^C,e_2,e_4),
	\end{equation}
	where $j=1,2,\dots,n_y$, and $n_y$ is the number of likely bit strings,
	
	\begin{equation}
		\begin{split}
			& n_y = 2^{2nH_2(E_y)}.\\
		\end{split}
	\end{equation}
	The function $f(Y^B\oplus Y^C,e_2,e_4)$ means that,
	based on the bit string $Y^B\oplus Y^C$ and the corresponding bit-flip error rate $E_y$, which has a maximum value of $(e_2+e_4)$, Bob can generate likely bit strings $\{K^{YB}_j\}$ relative to $Y^A$, i.e., $Y^A\in \{K_j^{YB}\}$.
	Of course, the bit-flip error rates $E_x$ and $E_y$ must satisfy $E_x,\ E_y \in(0,0.5)$.
	
	After generating the likely bit strings locally,
	Bob decrypts the signature $S$ with likely bit strings $\{K_j^{YB}\}$ to obtain $n_y$ expected digests,
	
	\begin{equation}\label{equ:Aed}
		S \oplus \{K_{j}^{YB}\} = \{(h_{j}^B,p_{j}^B)\},
	\end{equation}
	where $j = 1,2,\dots,n_y$.
	
	Bob then judges whether the expected hash values received from Alice have been generated from one likely bit string
	according to the expected hash values he received and his own raw keys. Bob can do this job in the following way, for example:
	Bob generates $n_xn_y$ LFSR-based Toeplitz matrices~\cite{krawczyk1994lfsr} $\{H_{i,j}^B\}$ based on likely bit strings $\{K_i^{XB}\}$ and $\{p_j^B\}$.
	After that, Bob performs $n_xn_y$ LFSR-based Toeplitz matrices $\{H_{i,j}^B\}$ on the message $M$ to obtain $n_xn_y$  actual hash values,
	\begin{equation}
		\{H^{B}_{i,j}\}\cdot M=\{\mathbb{h}^B_{i,j}\},
	\end{equation}
	which together with $\{p_j^B\}$ form actual digests,
	
	\begin{equation}\label{equ:Aad}
		\{(\mathbb{h}^B_{i,j},p_j^B)\}.
	\end{equation}

	Finally, Bob compares expected digests $\{(h_j^B,p_j^B)\}$~\eqref{equ:Aed} with actual digests $\{(\mathbb{h}_{i,j}^B,p_j^B)\}$~\eqref{equ:Aad} using the following comparison function $\mathbb{C}^B$,
	\begin{equation}\label{equ_comB}
		\mathbb{C}^B = \left\{
		\begin{array}{l}
			1 \quad \exists\ C_{i,j}^B=1, \\
			0 \quad \mathrm{else},\\
		\end{array}
		\right.
	\end{equation}
	where $C^B_{i,j}$ is
	
	\begin{equation}
		C^B_{i,j} = \left\{
		\begin{array}{l}
			1 \quad h_j^B=\mathbb{h}_{i,j}^B, \\
			0 \quad \mathrm{else}.\\
		\end{array}
		\right.
	\end{equation} 
	Obviously, this comparison function implies that Bob makes at most $n_xn_y$ comparisons. If $\mathbb{C}_B=1$, Bob accepts the message $M$; otherwise, Bob rejects the message $M$.
	
	\subsubsection{Verification of Charlie}
	
	Charlie uses similar steps to generate the following two sets of likely bit strings $\{K^{XC}_i\}$ and $\{K^{YC}_j\}$,
	
	\begin{equation}
		\{K^{XC}_i\} = f(X^B\oplus X^C,e_1,e_3),
	\end{equation}
	
	\begin{equation}
		\{K^{YC}_j\} = f(Y^B\oplus Y^C,e_2,e_4),
	\end{equation}
	where $i=1,2,\dots,n_x$ and $j=1,2,\dots,n_y.$
	
	Then, Charlie decrypts the signature $S$ with $\{K_j^{YC}\}$ to obtain $n_y$ likely expected digests, 
	\begin{equation}\label{equ:Bed}
		S \oplus \{K_{j}^{YC}\}=\{(h_{j}^C,p_{j}^C)\}.
	\end{equation}
	
	Charlie then judges whether the expected hash values received from Bob have been generated from one likely bit string
	according to the expected hash values he received and his own raw keys. Charlie can do this job in the following way, for example:
	Based on likely bit strings $\{K_i^{XC}\}$ and $\{p_j^C\}$, Charlie can obtain $n_xn_y$ LFSR-based Toeplitz matrices~\cite{krawczyk1994lfsr} $\{H_{i,j}^C\}$.
	After that, Charlie performs $\{H_{i,j}^C\}$ on the message $M$ to obtain $n_xn_y$ actual hash values, 
	
	\begin{equation}
		\{H^{C}_{i,j}\}\cdot M=\{\mathbb{h}^C_{i,j}\},
	\end{equation}
	which together with $\{p_j^C\}$ form the actual digests,
	
	\begin{equation}\label{equ:Bad}
		\{(\mathbb{h}^C_{i,j},p_j^C)\}.
	\end{equation}

	Finally, Charlie compares expected digests $\{(h_j^C,p_j^C)\}$~\eqref{equ:Bed} with actual digests $\{(\mathbb{h}_{i,j}^C,p_j^C)\}$~\eqref{equ:Bad} using the following comparison function $\mathbb{C}^C$,
	\begin{equation}\label{equ_com}
		\mathbb{C}^C = \left\{
		\begin{array}{l}
			1 \quad \exists\ C_{i,j}^C=1, \\
			0 \quad \mathrm{else},\\
		\end{array}
		\right.
	\end{equation}
	where $C^C_{i,j}$ is
	\begin{equation}
		C^C_{i,j} = \left\{
		\begin{array}{l}
			1 \quad h_j^C=\mathbb{h}_{i,j}^C, \\
			0 \quad \mathrm{else}.\\
		\end{array}
		\right.
	\end{equation}
	Obviously, this comparison function implies that Charlie makes at most $n_xn_y$ comparisons. 
	If $\mathbb{C}_C=1$, Charlie accepts the message $M$; otherwise, Charlie rejects the message $M$. 
	
	%%%%%%%%%%%%%%%%%%%%%%%%%%%%%%%%%%%%%%%%%%%%%%%%%%%%%%%%%%%%%%%%%%%%%%%%%%%%%%%%%%%%%
	
	\section{Security Analysis}\label{sec3}
	
	In the QDS protocol with three participants, at most one participant is malicious due to the majority principle. We analyze the security of our QDS protocol in terms of forgery attacks, repudiation attacks, and robustness~\cite{amiri2016secure,Yin2022QDShash}.
	
	First, we consider that Bob is malicious, i.e., he wants to launch a forgery attack. A successful forgery of Bob means that he can make Charlie accept a tampered message $M^{\prime}$, where $M^{\prime} \neq M$. In our protocol, Bob needs to send the message and the corresponding signature $(M,S)$ to Charlie before Charlie sends $X^C$ and $Y^C$ to Bob. Therefore, if Bob successfully guesses $X^C$ and $Y^C$, then combined with $X^B$ and $Y^B$, he can tamper with the message $M$ for Charlie to accept $M^{\prime}$. The probability of successfully guessing the bit string $X^C$ and $Y^C$ in Charlie's hand is
	
	\begin{equation}
		p_g \le 2^{-n\cdot H_2(p_e)},
	\end{equation}
	where $p_e$ is the minimum rate at which Bob can make errors when guessing $X^C$ and satisfies
	\begin{equation}\label{equ_pe}
		H_2(p_e)=\Delta_{1}[1-H_2(e^{ph})],
	\end{equation}
	in which $H_2(.)$ is the binary Shannon entropy, and  $\Delta_1$ and $e^{ph}$ are the proportion of effective bits of a single-photon state in $X^C$ and the phase-flip error rate of $X^C$, respectively. We define the bits generated in time windows of KGP where the clicks of detectors satisfy certain conditions as effective bits. For example, in SNS KGP~\cite{wang2018twin}, a bit generated in a time window with only one detector clicks is an effective bit.
	
	In addition to guessing the bit strings of Charlie, Bob has another method of forgery, which is to directly construct a set of tampered messages and corresponding signatures $(M^{\prime},S^{\prime})$ for Charlie to accept, where $M^{\prime}\neq M$ and $S^{\prime}\neq S$ or $S^{\prime}= S$. Because of the collision probability of LFSR-based Toeplitz matrices and the characteristics of our comparison function, Charlie may accept the tampered message $M^{\prime}$ with a very small probability. The collision refers to the situation where different messages correspond to the same hash value. The collision probability of LFSR-based Toeplitz matrices is at most $\frac{m}{2^{n-1}}$~\cite{krawczyk1994lfsr}, where $m$ is the length of the message and $n$ is the degree of irreducible polynomials used to generate the LFSR-based Toeplitz matrices. In our protocol, Charlie decides whether to accept the message based on the comparison function $\mathbb{C}_C$~\eqref{equ_com}, which compares the expected digests with the actual digests at most $n_xn_y$ times. For one fixed set of $(i, j)$, the upper bound of the probability that the expected hash value $h_j^{C,M^{\prime}}$ and actual hash value $\mathbb{h}_{i,j}^{C,M^{\prime}}$ of $M^{\prime}$ satisfy $h_j^{C,M^{\prime}}=\mathbb{h}_{i,j}^{C,M^{\prime}}$ is $\frac{m}{2^{n-1}}$. For other $(n_xn_y-1)$ sets of $(i, j)$, the upper bound of the probability that the hash values of $M^{\prime}$ satisfy $h_j^{C,M^{\prime}}=\mathbb{h}_{i,j}^{C,M^{\prime}}$ is $(\frac{m}{2^{n-1}})^2$. According to our comparison function, Charlie will reject $M^{\prime}$ only if the results of all $n_xn_y$ comparisons are not equal. As long as there is one equal comparison result, Charlie will accept $M^{\prime}$.
	Therefore, the probability that Charlie accepts $(M^{\prime},S^{\prime})$ is
	
	\begin{equation}
		p_h \le 1-(1-\frac{m}{2^{n-1}})[1-(\frac{m}{2^{n-1}})^2]^{n_xn_y-1}.
	\end{equation}
	Combining the above two scenarios, the probability of Bob's successful forgery is
	
	\begin{equation}\label{eq:pf}
		p_f = \{p_g, p_h\}_{\max}.
	\end{equation}
	
	Next, we consider that Alice is malicious, i.e., she wants to deny her signature. If Alice can make Bob accept the message $M$ and Charlie reject $M$, then she can deny the signed message, which is a successful repudiation attack. In our protocol, when Bob and Charlie are both honest, they generate the same possible bit strings locally, i.e., $\{K_i^{XB}\}=\{K_i^{XC}\}$  and $\{K_j^{YB}\}=\{K_j^{YC}\}$. Then, they generate the same LFSR-based Toeplitz matrices locally as well. Therefore, if Bob accepts the message $M$, then Charlie must also accept $M$. So, the probability of successful repudiation is
	
	\begin{equation}\label{eq:pre}
		p_{re} = 0.
	\end{equation}
	
	Finally, we need to analyze the robustness of our QDS protocol, i.e., the case where Bob rejects the message $M$ if all participants are honest. In this case, the likely bit strings $\{K_i^{XB}\}$ and $\{K_j^{YB}\}$ generated by Bob must contain $X^A$ and $Y^A$, respectively, i.e., $X^A\in \{K_i^{XB}\}$ and $Y^A\in \{K_j^{YB}\}$. According to our comparison function $\mathbb{C}_B$~\eqref{equ_comB}, Bob must accept the message $M$. Therefore, the probability that Bob rejects the message when all participants are honest is
	
	\begin{equation}\label{eq:pro}
		p_{ro} = 0.
	\end{equation}
	
	Based on the above analysis, we can define $\varepsilon$ as the security level~\cite{amiri2016secure} of our QDS protocol:
	
	\begin{Def}($\varepsilon$-secure). 
		If $\varepsilon$ satisfies 
		\begin{equation}\label{Eq:sl}
			\varepsilon=\{p_f,p_{re},p_{ro}\}_{\max},
		\end{equation}
		then we say that the security level of the QDS protocol is $\varepsilon$, or that the QDS protocol is $\varepsilon$-secure.
		
	\end{Def}

	%%%%%%%%%%%%%%%%%%%%%%%%%%%%%%%%%%%%%%%%%%%%%%%%%%%%%%%%%%%%%%%%%%%%%%%%%%%%%%%%%%%%%
	
	\section{Improved method}\label{sec4}
	In our protocol, we have given an example of the way for receivers to verify that the expected hash values received from Alice have been generated from one likely bit string, by listing all possible hash values from likely strings. In order to save computing resources, here we propose an improved method where Alice is involved in the verification process. The improved method does not require large complex computing relating to the huge number of likely strings while keeping the good performance of QDS in terms of the signature rate and distance without consuming large computing resources.
	
	\subsection{Improved protocol}
	
	In the signing stage of the protocol, Alice follows the steps described in Sec.~\ref{sec2} to generate the signature $S$ of message $M$.
	At the end of the signing stage, Alice sends $(M,S)$ to a receiver for verification. 
	
	In the verification stage, Bob forwards $(M,S)$ received from Alice to Charlie and also sends the bit strings $(X^B,Y^B)$ to Charlie. Then Bob informs Alice that he has received $(M,S)$. Next, Charlie sends the bit strings $(X^C,Y^C)$ to Bob and informs Alice that he has received $(M,S)$. Alice separately verifies whether both Bob and Charlie have received $(M,S)$. If both Bob and Charlie have received $(M,S)$, then Alice publishes $(X^A,Y^A)$; otherwise, the protocol is terminated. 
	
	Bob verifies the message $M$ based on the local bit strings and the information published by Alice. Bob obtains the expected digest $D^E_B$ using $Y^A$ and $S$,
	\begin{equation}
		D^E_B=(h^B,p^B)=S\oplus Y^A.
	\end{equation}
	Next, based on $X^A$ and $p^B$, Bob generates the LFSR-based Toeplitz matrix $H^B$ and acts it on the message $M$ to obtain the actual hash value
	\begin{equation}
		H^{B}\cdot M=\mathbb{h}^B.
	\end{equation}
	If $\mathbb{h}^B = h^B$, Bob accepts the message $M$; otherwise, Bob rejects the message $M$. 
	
	Charlie uses similar steps for verification. Charlie obtains the expected digest $D^E_C$ using $Y^A$ and $S$,
	\begin{equation}
		D^E_C=(h^C,p^C)= S\oplus Y^A.
	\end{equation}
	Next, based on $X^A$ and $p^C$, Charlie generates the LFSR-based Toeplitz matrix $H^C$ and acts it on the message $M$ to obtain the actual hash value
	\begin{equation}
		H^{C}\cdot M=\mathbb{h}^C.
	\end{equation}
	If $\mathbb{h}^C = h^C$, Charlie accepts the message $M$; otherwise, Charlie rejects the message $M$. 
	
	\subsection{Security analysis of improved protocol}
	
	We can perform a similar security analysis of the improved protocol in terms of unforgeability, nonrepudiation and robustness using the methods in Sec.~\ref{sec3}.  
	In our improved protocol, Alice will not publish $(X^A,Y^A)$ until she confirms that both Bob and Charlie have received $(M,S)$. That is, Bob does not know the bit string $(X^A,Y^A)$ before he sends the message to Charlie. Therefore, for Bob, the probability that he makes Charlie accept the forged message is still $p_f$ \eqref{eq:pf}, i.e., either due to successfully guessing Charlie's local bit string $X^C$, $Y^C$ or due to the collision probability of LFSR-based Toeplitz matrices. In terms of nonrepudiation, since both Bob and Charlie in the improved protocol use $(X^A,Y^A)$ published by Alice in their verification stage, their results of verification are consistent. Therefore, the probability of successful repudiation is still $p_{re}$ \eqref{eq:pre}. In terms of robustness, when all participants are honest, Alice will publish bit $(X^A,Y^A)$ after she has confirmed that both Bob and Charlie have received $(M,S)$. Then, Bob will obtain $\mathbb{h}^B = h^B$ using the bit string published by Alice. Thus Bob will accept the message $M$ when all participants are honest, i.e., the probability associated with robustness is still $p_{ro}$ \eqref{eq:pro}. 
	In general, the expression for the security level $\varepsilon$ of the improved protocol is the same as that \eqref{Eq:sl} of the protocol introduced in Sec.\ref{sec2}.
	
	Compared with our protocol in Sec.~\ref{sec2}, our improved protocol can significantly reduce the computing resources without compromising security and efficiency. In our improved protocol, instead of using the old verification method, which may need large computing resources, such as listing all possible hash values from likely strings, Bob and Charlie utilize the bit strings published by Alice for verification, which completely eliminates the need for large computing resources.  
	
	\section{Discussion}
	
	Our protocol can be applied to any QKD-based QDS protocol. Using existing mature QKD devices, based on the idea of likely bit strings, we can implement efficient QDS over longer distances. We use the signature rate~\cite{zhang2021twin,qin2022quantum} $R$ to describe the efficiency of the QDS protocol,
	
	\begin{equation}\label{Equ:SR}
		R = \frac{m}{2N},
	\end{equation}
	where $m$ is the length of the message $M$ and $N$ is the total number of pulses in the KGP between Alice and Bob, i.e., the whole process of generating $X_A^B$ and $Y_A^B$. The total number of pulses in the KGP between Alice and Charlie, i.e., the whole process of generating $X_A^C$ and $Y_A^C$, in the symmetric case is also $N$. Therefore, Eq.~\eqref{Equ:SR} has a factor of $2$ in the denominator.
	
	We demonstrate the advantages of our method and improved method using QDS with SNS KGP~\cite{wang2018twin} as an example. Alice performs the quantum communication of SNS QKD, i.e., SNS KGP, with Bob and Charlie, respectively, to generate the associated bit strings. After the associated bit strings are generated among the participants, the signing and verification steps can be performed as described in Sec.~\ref{sec2} and Sec.~\ref{sec4}. 
	In practical applications of QDS, the total number of pulses is finite. We need to estimate the parameters while considering the effects of finite data size.
	In our QDS protocol as well as security analysis, we need to obtain the bit-flip error rate $E$, the proportion of effective bits  of a single-photon state $\Delta_1$, and the phase-flip error rate $e^{ph}$ of the associated bit strings. In the Appendix~\ref{Appendix_SNS}, we describe in detail how to obtain these parameters.	
	
\begin{figure}
		\centering
		\includegraphics[scale=0.6]{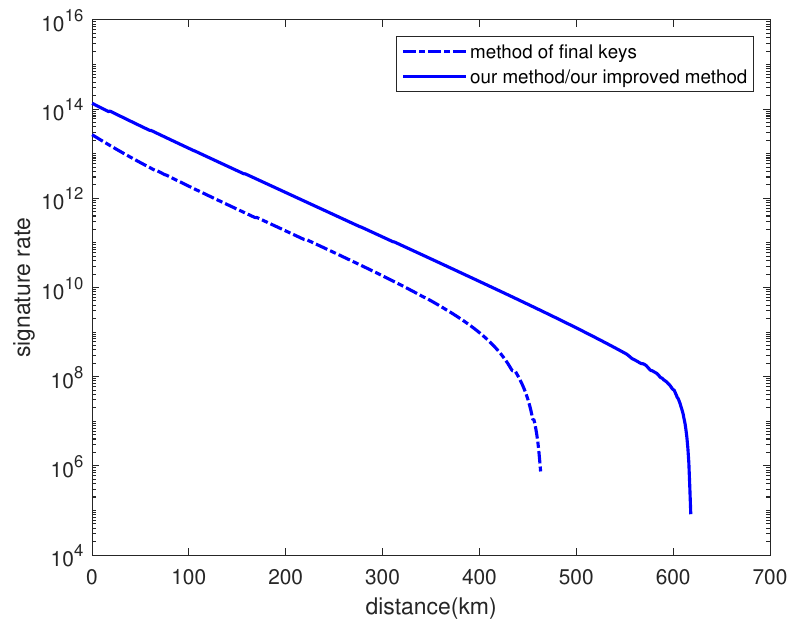}
		\caption{Comparison of the signature rate with our method, our improved method and the method of final keys. The solid line represents the signature rate of our method and our improved method, and the broken line represents the signature rate using the method of final keys. Under the system parameters shown in Table~\ref{table_1} and the length of the message $m=10^{20}$, both our method and our improved method can improve the signature rate by more than $100$ times, and increase the secure distance of QDS by about $150$ km.}\label{picture_2}
	\end{figure}

	In Fig.~\ref{picture_2}, considering the effects of finite data size, we compare the signature rate with our method, our improved method and the method using final keys. The solid line indicates the signature rate obtained by our method and our improved method, and the broken line indicates the signature rate obtained by the method of final keys. As can be seen in Fig.~\ref{picture_2}, both our method and our improved method can significantly improve the signature rate and dramatically increase the secure distance of QDS.
	The signature rate can be improved by more than $100$ times, and the secure distance can be increased by about $150$ km using our method or our improved method under the system parameters in Table~\ref{table_1} and the length of the message $m=10^{20}$.
	It should be emphasized that, as introduced in Sec.~\ref{sec4}, the improved method can improve the performance of QDS in terms of signature rate and distance as well as save computing resources.  
	
	\begin{table}[]
		\vspace{20pt}
		\centering
		\caption{System parameters used in our numerical simulation. Notation: $\alpha$ is the loss coefficient of the fiber; $e_d$ is the misalignment error rate; $p_d$ is the dark count rate of the detector; $\eta_d$ is the efficiency of the detector; $\xi$ and $\varepsilon_p$ are failure probabilities in parameter estimation, which are described in detail in the Appendix~\ref{Appendix_SNS}; and $\varepsilon$ is the security level of our QDS protocol.}
		\begin{tabular}{p{1.6cm}p{0.9cm}p{0.9cm}p{0.9cm}p{0.9cm}p{0.9cm}p{0.9cm}}
			\hline
			\hline
			$ \alpha $  & $ e_d $ & $ p_d $ & $\eta_d$ & $\xi$ & $\varepsilon_p $& $\varepsilon$  \\
			\hline
			$ 0.2\ \mathrm{dB/km} $ & $ 2\%$ &$ 10^{-8} $ & $50\% $ & $10^{-12}$ & $10^{-10}$ & $10^{-10}$ \\
			\hline
			\hline
		\end{tabular}
		\label{table_1}
	\end{table}
		
	During the implementation of our protocol, the actual number of comparisons $n_0$ that the receiver needs to perform satisfies $n_0\le n_xn_y$. This is because, according to the characteristics of our comparison functions $\mathbb{C}^B$~\eqref{equ_comB} and $\mathbb{C}^C$~\eqref{equ_com}, as soon as an equal set of expected and actual digests is found, the receiver can stop the subsequent comparisons and accept the message $M$.
	
	In this work, we use LFSR Toeplitz hash functions~\cite{krawczyk1994lfsr} in our QDS protocol. Of course, our method of using likely bit strings for QDS can be extended to the case of using other types of hash functions. For protocols that use other types of hash functions, the corresponding security analysis can be performed using similar ideas based on the collision probability of hash functions as in this work.
	
	In order to reduce the consumption of computing resources, we propose an improved protocol based on our protocol in Sec.~\ref{sec2}. In our improved protocol, Alice is involved in the verification process of Bob and Charlie. On the one hand, Alice publishes the bit strings $(X^A,Y^A)$ after confirming that both Bob and Charlie have received $(M,S)$. Therefore, this does not pose a threat to security. On the other hand, Bob and Charlie can directly use $(X^A,Y^A)$ published by Alice for verification, which no longer need to consume large computing resources.
	
	In conclusion, based on the idea of likely bit strings, we propose an efficient and practicable QDS protocol over long distances. Our method only requires mature QKD devices, while improving the performance of QDS in terms of signature rate and distance. With typical parameters, our method of likely bit strings can improve the signature rate by more than $100$ times and increase the distance of QDS by about $150$ km compared with a hash function-based QDS protocol using final keys. To save computing resources, we propose an improved method, which can significantly improve the signature rate and distance of QDS without large computing resources.

	%{\bf{Acknowledgements:}} 
	\begin{acknowledgements}
		We acknowledge partial financial support by the National Natural Science Foundation of China, Grants No.12174215,  No.12374473, and No.11974204; and by the Taishan Scholars Program.
		%; Open Research Fund Program of the State Key Laboratory of Low-Dimensional Quantum Physics No.KF202110.
	\end{acknowledgements}
	
	\appendix
	
	\section{Sending-or-Not-Sending Key Generation Protocol}
	\label{Appendix_SNS}
	
	We take the QDS with SNS KGP as an example to demonstrate the advantages of our method of likely bit strings. We generate the associated bit strings by the quantum communication of SNS QKD~\cite{wang2018twin,Yu2019SNSdecoy,jiang2019unconditional}.
	
	In each time window, Alice (Bob) chooses the window as a signal window or decoy window with probability $p_z$ or $(1-p_z)$. In the signal window, Alice (Bob) writes down the bit value $1$ ($0$) when sending the phase-randomized coherent state of intensity $\mu$ to Eve with probability $q$ and writes down the bit value $0$ ($1$) when not sending the coherent state, i.e., sending the vacuum state, to Eve with probability $(1-q)$. In the decoy window, Alice (Bob) randomly sends phase-randomized coherent states of intensities $\mu_0=0$, $\mu_1$, and $\mu_2$, with probability $p_0$, $p_1$, and $(1-p_0-p_1)$, to Eve. Eve performs measurements using beam splitters and announces the outcome of measurements. We can record the event where one and only one detector clicks as an effective event. The corresponding time window of an effective event is an effective window and the corresponding bit is an effective bit. The effective bits used for parameter estimation are discarded. And then, Alice and Bob obtain the associated bit strings of raw keys.
	
	The receivers can construct likely bit strings based on the bit strings of raw keys and the bit-flip error rates and use these likely bit strings for verification. In the following, we show how to obtain the bit-flip error rate $E$, the proportion of effective bits  of a single-photon state $\Delta_1$, and the phase-flip error rate $e^{ph}$ of the associated bit strings taking into account the effects of finite data size. 
	
	Effective events can be divided into three categories with counting rates $S_C$, $S_D$ and $S_V$, respectively, with
	\begin{equation}\label{}
		\begin{split}
			&S_{C}=2[(1-p_d)e^{-\eta \mu/2}-(1-p_d)^2e^{-\eta\mu}],\\
			&S_{D}=2[(1-p_d)e^{-\eta\mu}I_0(\eta\mu)-(1-p_d)^2e^{-2\eta\mu}],\\
			&S_{V}=2p_d(1-p_d),\\
		\end{split}
	\end{equation}
	where $S_C$ corresponds to the case where one and only one participant chooses to send and the other participant chooses not to send, and $S_D$ and $S_V$ represent the case where both participants choose to send and both participants choose not to send, respectively. The notation $p_d$ represents the dark count rate of detectors, $\eta=10^{-\frac{\alpha l}{10}}\eta_d$ is the total system efficiency, $\alpha$ is the loss coefficient of the fiber, $l$ is half of the distance between the two participants (we assume that the node used for measurement in QDS with SNS KGP is midway between the two participants), $\eta_d$ represents the detection efficiency, $\mu$ is the intensity of the coherent state and $I_0(x)$ is the $0$-order hyperbolic Bessel function of the first kind.
	
	Based on the above three counting rates $S_C$, $S_D$ and $S_V$, we can obtain their effective event numbers $n_C$, $n_D$, and $n_V$ as follows:
	
	\begin{equation}
		\begin{split}
			& n_{C}=2Np_z^2q(1-q)S_{C},\\
			& n_{D}=Np_z^2q^2S_{D},\\
			& n_{V}=Np_z^2(1-q)^2S_{V}.\\
		\end{split}
	\end{equation}
	Here $N$ is the total number of pluses in SNS KGP between two participants, $p_z$ is the probability that the participant chooses the time window as a signal window (i.e., the participant chooses the time window as a decoy window with probability $(1-p_z)$) and $q$ represents the probability that the participant chooses to send.

	Therefore, the bit-flip error rate $E_T$ of the bit strings generated by SNS KGP is
	\begin{equation}\label{equ_E}
		\begin{split}
			&E_T = \frac{n_{D}+n_{V}}{{N_{t}}},\\
		\end{split}
	\end{equation}
	where $N_t$ is the total number of three effective events
	\begin{equation}
		\begin{split}
			& N_t=n_{C}+n_{D}+n_{V}.\\
		\end{split}
	\end{equation}
	
	It should be noted that, in experiments, the bit-flip error rate $E_T$ can be directly observed. In our numerical simulations, we use the expression \eqref{equ_E} of $E_T$. We randomly select $T$ (in our simulation, we set $T = \gamma N_t$ and $\gamma=10\%$) bits to obtain their bit-flip error rate, which is used to estimate the bit-flip error rate $E$ of the remaining bits~\cite{Serfling1974,amiri2016secure,puthoor2016measurement},  
	
	\begin{equation}\label{equ_Er}
		\begin{split}
			&E \le E_T+\mu(n,T,\varepsilon_{p}),\\
		\end{split}
	\end{equation}
	with
	\begin{comment}
		\begin{equation}
			\begin{split}
				&   \mu(n,T,\varepsilon_{\mathrm{PE}})=\frac{1}{n}\sqrt{\frac{(n+1)(n+T)\ln(\frac{1}{\varepsilon_{\mathrm{PE}}})}{2T}},\\
			\end{split}
		\end{equation}
	\end{comment}
	
	\begin{equation}
		\begin{split}
			&   \mu(n,T,\varepsilon_{p})=\sqrt{\frac{(n-T+1)\ln(\frac{1}{\varepsilon_{p}})}{2nT}},\\
		\end{split}
	\end{equation}
	where $\varepsilon_p$ is the failure probability of this estimation, and $n$ is the length of $X^A$,
	
	\begin{equation}
		n = \lfloor \frac{N_t-T}{3} \rfloor.
	\end{equation}
	
	Next, we analyze how to obtain the proportion $\Delta_{1}$ of effective bits  of a single-photon state and phase-flip error rate $e^{ph}$ of the bit string $X^C$, which are used in the calculation of $p_e$~\eqref{equ_pe}.
	
	We use the Chernoff bound~\cite{chernoff} to estimate the expected values $\phi$ from the observed values, which can be obtained directly in experiments, and then we use the Chernoff bound to estimate the real values $\varphi$ of a specific experiment from the expected values. The lower and upper bounds of the expected value are denoted by $\phi^{L}$ and $\phi^{U}$, respectively,
	
	\begin{equation}
		\begin{split}
			&\phi^{L}(X)=\frac{X}{1+\delta_1(X)},\\
			&\phi^{U}(X)=\frac{X}{1-\delta_2(X)},\\
		\end{split}
	\end{equation}	
	and $\delta_1(X)$ and $\delta_2(X)$ satisfy
	
	\begin{equation}
		\begin{split}
			&(\frac{e^{\delta_1}}{(1+\delta_1)^{1+\delta_1}})^{\frac{X}{1+\delta_1}}=\frac{\xi}{2},\\
			&(\frac{e^{-\delta_2}}{(1-\delta_2)^{1-\delta_2}})^{\frac{X}{1-\delta_2}}=\frac{\xi}{2},\\
		\end{split}
	\end{equation}	
	where $\xi$ denotes the failure probability of estimation.
	
	The lower and upper bounds of the real value are denoted by $\varphi^{L}$ and $\varphi^{U}$, respectively,
	\begin{equation}
		\begin{split}
			&\varphi^{L}(Y)=[1+\delta_1^{\prime}(Y)]Y,\\
			&\varphi^{U}(Y)=[1-\delta_2^{\prime}(Y)]Y,\\
		\end{split}
	\end{equation}	
	and $\delta_1^{\prime}(X)$ and $\delta_2^{\prime}(X)$ satisfy
	\begin{equation}
		\begin{split}
			&(\frac{e^{\delta_1^{\prime}}}{(1+\delta_1^{\prime})^{1+\delta_1^{\prime}}})^{Y}=\frac{\xi}{2},\\
			&(\frac{e^{-\delta_2^{\prime}}}{(1-\delta_2^{\prime})^{1-\delta_2^{\prime}}})^{Y}=\frac{\xi}{2},\\
		\end{split}
	\end{equation}	
	where $\xi$ represents the failure probability of estimation. 
	
	Based on the analysis in SNS QKD~\cite{wang2018twin,Yu2019SNSdecoy,jiang2019unconditional}, we can obtain the lower bound of the expected value of the counting rate of single-photon state $\langle s_{1}^{\mathrm{L}} \rangle$ and the upper bound of the expected value of the phase-flip error rate $\langle e^{\mathrm{ph},\mathrm{U}} \rangle$,
	
	\begin{equation}\label{equ_s1}
		\langle s_1\rangle \geq \langle s_{1}^{L} \rangle=\frac{1}{2}(\langle s^{L}_{01}\rangle+\langle s^{L}_{10}\rangle),
	\end{equation}
	where $\langle s_{01}^{{L}}\rangle$ and $\langle s_{10}^{{L}}\rangle$ satisfy 
	\begin{equation}\label{}
		\begin{split}
			&\langle s_{01}^{{L}}\rangle=\frac{\mu_2^2e^{\mu_1}\langle S_{01}^{{L}}\rangle-\mu_1^2e^{\mu_2}\langle S_{02}^{{U}}\rangle-(\mu_2^2-\mu_1^2)\langle S^{{U}}_{00}\rangle}{\mu_1\mu_2(\mu_2-\mu_1)},\\
			&\langle s_{10}^{{L}}\rangle=\frac{\mu_2^2e^{\mu_1}\langle S_{10}^{{L}}\rangle-\mu_1^2e^{\mu_2}\langle S_{20}^{{U}}\rangle-(\mu_2^2-\mu_1^2)\langle S^{{U}}_{00}\rangle}{\mu_1\mu_2(\mu_2-\mu_1)},\\
		\end{split}
	\end{equation}
	and
	\begin{equation}\label{equ_eph}
		\langle e^{{ph}} \rangle\le \langle e^{{ph},{U}} \rangle=\frac{\langle T_{\Delta}^{{U}}\rangle-\frac{1}{2}e^{-2\mu_1}\langle S^{{L}}_{00}\rangle}{2\mu_1e^{-2\mu_1}\langle s_1^{{L}}\rangle}.
	\end{equation}
	The notation $S_{ij}$, $i,j
	\in\{0,1,2\}$, in the above equations denotes the counting rate of source $\mu_i\mu_j$ when participants choose coherent state $\mu_i$ and $\mu_j$ to send.  
	In the decoy window, participants randomly select three different intensities $\mu_0=0,\ \mu_1,\ \mu_2$ of phase-randomized coherent states with probability $p_0,\ p_1,\ (1-p_0-p_1)$, respectively. The notation $\langle T_{\Delta}^{{U}} \rangle$ can be calculated by
	
	\begin{equation}\label{}
		\langle T_{\Delta}\rangle\le \langle T_{\Delta}^{{U}} \rangle = \frac{\phi^{{U}}(n_{\Delta^+}^{l}+n_{\Delta^-}^{r})}{2N_{\Delta^{\pm}}},
	\end{equation}
	where $n_{\Delta^{+}}^{l}$ and $n_{\Delta^{-}}^{r}$ represent effective events heralded by the left and right detector, respectively,
	
	\begin{equation}
		n_{\Delta^{+}}^{l} =n_{\Delta^{-}}^{r}= [T_X(1-2e_d)+e_dS_X]N_{\Delta^{\pm}},
	\end{equation}
	\begin{equation}
		N_{\Delta^{\pm}}=\frac{\Delta}{2\pi}(1-p_z)^2p_1^2N,
	\end{equation}
	and
	\begin{equation}
		T_X=\frac{1}{\Delta}\int_{-\frac{\Delta}{2}}^{\frac{\Delta}{2}}(1-p_d)e^{-2\eta\mu_1\cos^2{\frac{\delta}{2}}}d\delta-(1-p_d)^2e^{-2\eta\mu_1},
	\end{equation}
	
	\begin{equation}
		\begin{split}
			S_X=&\frac{1}{\Delta}\int_{-\frac{\Delta}{2}}^{\frac{\Delta}{2}}(1-p_d)e^{-2\eta\mu_1\sin^2{\frac{\delta}{2}}}d\delta-(1-p_d)^2e^{-2\eta\mu_1}\\
			&+T_X.\\
		\end{split}
	\end{equation}
	We set $\Delta = \frac{\pi}{15}$ in our simulation.

	Based on $\langle s_{1}^{{L}} \rangle$~\eqref{equ_s1} and $\langle e^{{ph},{U}} \rangle$~\eqref{equ_eph}, we can use Chernoff bound to estimate the real values $\Delta_1$ and $e^{ph}$ from the expected values,
	
	\begin{equation}\label{equ_s1r}
		\Delta_1 = \frac{\varphi^{{L}}(n\langle \Delta_{1}^L \rangle)}{n}, 
	\end{equation}
	
	\begin{equation}\label{equ_ephr}
		e^{{ph}}=\frac{\varphi^{{U}}(n\Delta_{1}\langle e^{{ph},{U}}\rangle )}{n\Delta_{1}},
	\end{equation}
	where
	\begin{equation}
		\langle \Delta_{1}^{{L}} \rangle = \frac{2Np_z^2q(1-q)\mu e^{-\mu}\langle s_{1}^{{L}} \rangle}{N_t}.
	\end{equation}
	
	According to the bit-flip error rate $E$~\eqref{equ_Er}, the proportion of effective bits of a single-photon state $\Delta_1$~\eqref{equ_s1r}, and the phase-flip error rate $e^{ph}$~\eqref{equ_ephr}, we can optimize the signature rate $R$ by selecting the appropriate $(N,\mu,\mu_1,\mu_2,q,p_z,p_0,p_1)$  given security level $\varepsilon$ and other system parameters. 
	
	\bibliography{refs}

\end{document}